\documentclass[sigconf]{acmart}

\AtBeginDocument{%
  \providecommand\BibTeX{{%
    \normalfont B\kern-0.5em{\scshape i\kern-0.25em b}\kern-0.8em\TeX}}}

\usepackage{booktabs} % For professional looking tables
\usepackage{multirow} % Required for multirow cells
\usepackage{rotating} % Required for sidewaystable
\usepackage{caption}
\usepackage{comment}
\usepackage{xcolor}
\usepackage{soul}
\usepackage{balance}
\usepackage[skip=3pt]{caption}

\copyrightyear{2025}
\acmYear{2025}
\setcopyright{acmlicensed}
\acmConference[SIGCSE TS 2025] {Proceedings of the 56th ACM Technical Symposium on Computer Science Education V. 1}{February 26 -- March 1, 2025}{Pittsburgh, PA, USA.}
\acmBooktitle{Proceedings of the 56th ACM Technical Symposium on Computer Science Education V. 1 (SIGCSE TS 2025), February 26 -- March 1, 2025, Pittsburgh, PA, USA}
\acmISBN{979-8-4007-0531-1/25/2}
\acmDOI{10.1145/3641554.3701804}
\begin{document}

\title{Investigating the Presence and Development of Student Instructor Preferences in a Large-Scale CS1 Course}


\author{Yiqiu Zhou}
\affiliation{%
  \institution{University of Illinois Urbana-Champaign}
  \department{College of Education}
  \city{Champaign-Urbana}
  \state{IL}
  \country{USA}}
\email{yiqiuz3@illinois.edu}

\author{Luc Paquette}
\affiliation{%
  \institution{University of Illinois Urbana-Champaign}
  \department{College of Education}
  \city{Champaign-Urbana}
  \state{IL}
  \country{USA}}
\email{lpaq@illinois.edu}

\author{Geoffrey Challen}
\affiliation{%
  \institution{University of Illinois Urbana-Champaign}
  \department{College of Engineering}
  \city{Champaign-Urbana}
  \state{IL}
  \country{USA}}
\email{challen@illinois.edu}

\renewcommand{\shortauthors}{Anon et al.}

\begin{CCSXML}
<ccs2012>
   <concept>
       <concept_id>10010405.10010489.10010495</concept_id>
       <concept_desc>Applied computing~E-learning</concept_desc>
       <concept_significance>500</concept_significance>
       </concept>
   <concept>
       <concept_id>10010405.10010489.10010491</concept_id>
       <concept_desc>Applied computing~Interactive learning environments</concept_desc>
       <concept_significance>500</concept_significance>
       </concept>
 </ccs2012>
\end{CCSXML}

\ccsdesc[500]{Applied computing~E-learning}
\ccsdesc[500]{Applied computing~Interactive learning environments}
\keywords{Instructor Preference; Student-Faculty Relationship; Computer Science Education; Clickstream data; Log Analysis}


\begin{abstract}

% This study examines the presence and development of students’ instructor preferences over the duration of a CS1 course.
%
Prior research has established the importance of student instructor preferences and identified various influencing factors.
However, the dynamics of how student instructor preferences develop and change are less well understood, due to the limitations of common course structures and reliance on one-time measurements.
To bridge this gap, we utilize data from a novel learning platform that provides students with access to instructional content created by multiple instructors.
This platform enables the quantification of preference emergence and evolution throughout an entire semester, as students repeatedly select content from different instructors.
Examining both initial and final student instructor preferences suggests that preference is a dynamic construct continually shaped by experiences.
Furthermore, our analysis of the associations between preferences and student characteristics reveals a nuanced picture: while student attributes did not significantly correlate with initial preferences, substantial differences emerged in final preferences across genders and self-reported prior programming experience.
This analysis contributes to the existing body of knowledge by expanding our understanding of student instructor preferences and student-instructor relationships in computer science education. 
We also provide practical insights that institutions and instructors can draw on when multiple instructors collaborate on a course.
\end{abstract}

\maketitle

\section{Introduction}
Students prefer some instructors over others.
Student instructor preferences may be rooted in perception---one instructor may be perceived to be easier or more enjoyable based on an established reputation or feedback from peers.
Instructor subject knowledge, personality traits, communication skills, and teaching styles all play a role in shaping perception~\cite{barth2008deciphering, clayson2020student, phillips2017adult, seaton1980student,mcgoldrick2002instructor}, as do student characteristics, demographics, and preexisting biases~\cite{armstrong2007significance, criado2012quality, kim2009student}.
But instructor preferences may also reflect underlying realities---one instructor may actually be more effective than another, either for some students or for most.
Regardless of why, student perception of teaching impacts persistence, sense of belonging, academic motivation, educational aspiration, and learning~\cite{kim2009student,komarraju2010role,trolian2016influencing,trolian2017moderating}.
And overall student satisfaction influences key higher education outcomes: success, attrition, retention, and overall academic performance~\cite{duque2014framework,elliott2002student,mihanovic2016link,nastasic2019student,van2018first,wong2023student}.
When a student learns better from a preferred instructor, that is a clear educational benefit.
But even when a student learns equally well from several instructors, there are still benefits from allowing them to choose the one they prefer.

Common higher educational curriculum and course structures provide limited opportunities to study how student instructor preferences develop and evolve.
Students typically have few opportunities to even express instructor preferences: if a course is taught primarily by one instructor, or when they need to enroll in a course in a particular semester to graduate.
When multiple instructors do teach the same course in the same term, section scheduling or registration ordering may constrain student choices.
Limited information about the different instructors may be available to students, at least from reputable sources.
And the structure of typical multi-instructor courses means that, once a student makes their \textit{a priori} choice, they will have little or no contact with the other instructors.
As a result, they may have no way to directly compare instructors' core teaching competencies: the ability to effectively explain course content and motivate students to learn and succeed.
This environment provides no opportunity for dynamic instructor preference formation and evolution, and post-course evaluations will only reflect student satisfaction with their assigned instructor~\cite{mcgoldrick2002instructor}.

To better understand how student instructor preferences develop and change, we present analysis of data generated by a novel interactive online learning platform.
The platform sequences small content explanations from multiple instructors side-by-side, maintaining conceptual consistency while exposing students to different instructor voices and teaching styles.
The educational objective is to effectively integrate explanations from multiple instructors in an additive way---since students can freely view explanations from multiple instructors as needed.
But the repeated choices made by students as they become familiar with the instructors also produces fine-grained temporal data on instructor preference.
A student who began by sampling content from multiple instructors may begin to prefer one more exclusively, while another student who started out preferring one may have their preference change as the course continues.
Our dataset, drawn from a large CS1 course, provides a more comprehensive picture of how student instructor preferences develop and change compared with previous studies.
Below, we describe the platform's implementation, the process of creating course content, and present analysis showing how student instructor preferences form and evolve.
Additionally, we discuss the implications of our findings for platform design and CS education.

\raggedbottom

\section{Related Work}
Instructor preference is defined as a general term to describe students’ favorability of a specific instructor. This definition was adapted from literature on student satisfaction~\cite{elliott2002student, sirgy2010quality, weerasinghe2017students}, which examines overall satisfaction with college life in a broader sense. According to~\cite{sirgy2010quality}, an overarching framework to measure the quality of college life is composed of three aspects, including satisfaction with social, academic, and university facilities and services. Instructor preference specifically focuses on experiences and attitudes as a result of interactions with instructors. Therefore, instructor preferences can be conceptualized as students’ inclination towards a particular instructor based on subjective evaluations of the extent to which their expectations of learning experiences and academic outcomes have been fulfilled or exceeded.

Previous studies have mainly relied on surveys or interviews to explore the factors associated with students’ perceptions and preferences towards instructors~\cite{ahmad2015evaluating, clayson2020student}. These studies often assumed that preferences remain homogeneous over time. However,~\cite{elliott2002student} suggested that student satisfaction is shaped by repeated experiences. We hypothesize that instructor preference is similarly influenced by cumulative interactions. Thus, measuring instructor preferences at a single point may not accurately reflect students' evolving needs and favorability toward instructional approaches. For instance, some students may demonstrate stable preferences at an early stage, while others may take time to develop preferences and adjust them according to learning needs. 

Instructor preference is a complex and multidimensional construct, with extensive research highlighting numerous factors shaping its manifestation. These factors - shown in studies exploring students’ ratings of instructors~\cite{kardan2013prediction}, preferred characteristics of instructors~\cite{seaton1980student}, and student-faculty interactions~\cite{kim2009student} - can be broadly grouped into three categories: course, instructor, and student-related factors. Course-related attributes (e.g., curriculum design, course content, and classroom format) undoubtedly influence students' perceptions. However, faculty or instructor-level factors also impact the learning experiences and contribute to the overall instructor favorability. Studies have underscored the importance of faculty knowledge, quality of instruction, ability to fulfill students' learning requirements, and impartiality~\cite{elliott2002student, gopal2021impact}. Technological readiness and positive faculty reputation have also been shown to enhance student satisfaction~\cite{mihanovic2016link}.

Beyond pedagogical approaches and instruction quality, tangible factors, such as gender, also shape students' perceptions and preferences~\cite{macnell2015s}. For instance, research suggests that certain student groups exhibit gender-based preferences in elective course decisions~\cite{mcgoldrick2002instructor}. This has been further supported by studies demonstrating a persistent gender bias in student ratings of instructors~\cite{centra2000there, chavez2020exploring, macnell2015s, wong2021students, young2009evaluating}, with students often rating male instructors higher than their female counterparts, even when controlling for course materials and instruction quality.

Moreover, there is a need to consider student-related factors when examining preferred instructors. Literature indicates that students’ demographic profiles impact their perceptions and preferences regarding interactions with instructors. \cite{kim2009student} pointed out that student–faculty interaction preferred or disfavored by students can vary as a result of student gender, race, social class, and first-generation status. This view was echoed by~\cite{elliott2002student}, who argued for the need to recognize the student’s varying degrees of satisfaction with each factor relevant to overall satisfaction in university. These findings underscore that student-faculty interactions and their perceived benefits can significantly vary based on student gender and other subgroups.

Overall, previous studies have noted the presence of numerous factors associated with students' preferences for instructors. Most studies, however, rely on data from post-course evaluations and questionnaires, rather than probing the formation of these perceptions. As preferences can be shaped by both pre-existing inherent biases and various characteristics of instructors within courses~\cite{mcgoldrick2002instructor}, further research is needed to understand the evolution and development of these preferences over time.

\section{Learning Platform}
Our online CS1 platform provides a flexible and personalized learning experience.
This section details the course structure, platform design, and content creation process, highlighting the key innovation of allowing students to select content from multiple instructors at a fine granularity throughout each lesson.

\subsection{Course Structure}

Since Fall~2020, our CS1 course has transitioned from synchronous lectures to an asynchronous online tutorial format.
Students complete a series of online daily lessons combining text; code examples; interactive content, including both code walkthroughs and video explanations; practice programming exercises; and graded programming, testing, and debugging exercises.
Each week students complete a proctored computer-based quiz comprising multiple-choice questions on course concepts, programming tasks, and debugging challenges.
During the second half of the semester, students complete a longer Android programming project.
Students can choose to complete the course in either Kotlin or Java.
This format has proven successful: typically around 80\% of students earn A-range grades and less than 4\% fail.

\subsection{Platform Design}

A key design goal of our learning platform is incorporating content from multiple instructors---including faculty and graduate and undergraduate course staff---while also maintaining coherence and consistency.
We want to allow students to express their preferences and benefit from contributions from multiple instructors, but also ensure that they all learn the same material.

\begin{figure}
  \centering
  \includegraphics[width=\linewidth]{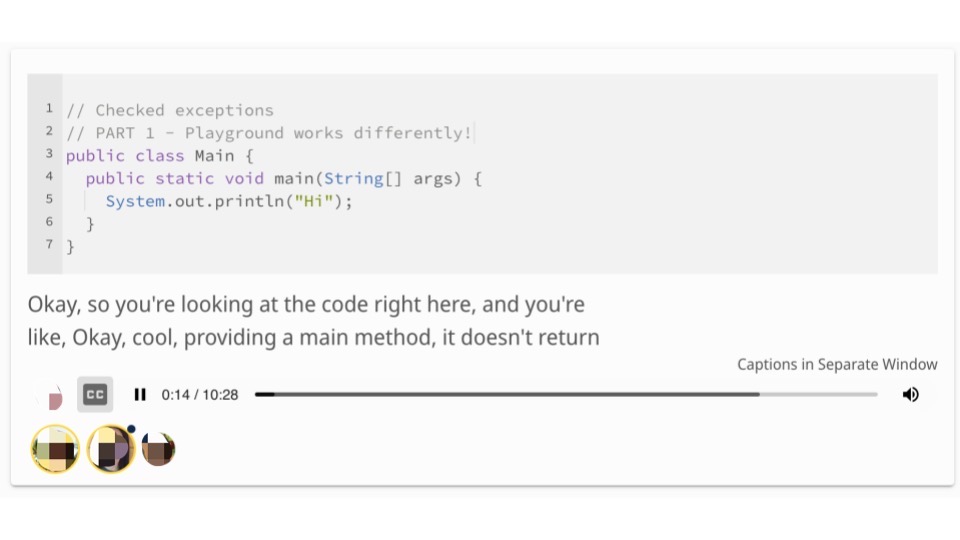}
  \caption{Screenshot of an interactive walkthrough in the CS1 learning platform. (Staff photos are anonymized.)}
  \label{fig:interface}
\end{figure}

Each daily lesson includes multiple interactive walkthroughs and video explanations for students to review.
Students select an instructor and begin playback by clicking on their profile photo displayed below the explanation player, as shown in Fig.~\ref{fig:interface}.
When multiple explanations are available for a given concept, faculty explanations are listed first, followed by course staff.
But students are free to watch any available explanations in any order.

Whenever possible, each interactive explanation explains a single new concept or bit of coding syntax in just a few minutes---for example, carefully explaining each component of a variable declaration.
Keeping content short makes it easier for students to repeat explanations they don't understand or review specific concepts later.
This design also makes it easier for instructors and staff to add new explanations or improve existing ones.
Short and focused explanations also help establish consistency across multiple instructors, ensuring that students learn the same concepts regardless of which explanations they review.

When multiple faculty explanations are available for a given concept, one faculty is chosen at random to be presented first.
This random assignment is stable across page reloads—so the same student sees the same faculty instructor first each time they return to that lesson.
In addition, the same random faculty instructor is shown first for all explanations across the entire lesson, to accommodate cases where multiple explanations form a sequence and mixing instructors would reduce coherence.

% The platform also makes it straightforward for course staff to add new explanations.
%
% Interactive code walkthroughs are recorded in the browser on the lesson page exactly where the explanation will be viewed by students, to ensure authors respond to the surrounding lesson context.
%
%Video explanations are first uploaded to YouTube and then added at the appropriate part of the page.

\subsection{Content Creation}

With multiple instructors contributing content, we implemented a structured creation process to ensure consistency.
Detailed guidelines are visible to instructors outlining the key concepts and examples to be covered in each code walkthrough or video explanation.
For instance, during a lesson on loops, one explanation instructs contributors to demonstrate the functionality of the while loop, while a second asks them to illustrate the difference in variable scope between the while and for loop.
Code walkthroughs are recorded in the browser on the lesson page exactly where the explanation will be viewed by students, to ensure authors understand the surrounding lesson context.
Video explanations are first uploaded to YouTube and then added at the appropriate part of the page.
Instructors also have access to existing content for review, to further reinforce the pedagogical purpose of each explanation.
This process ensures consistency while allowing instructors to leverage their distinct teaching styles and explanations.

\section{Evaluation Results}
%%\input{main/outline}

%%\section{Discussion}
Throughout the implementation of this multi-instructor CS1 platform, we collected data to inform potential improvements and evaluate whether students show preference for instructors.
This section presents key insights from this data, focusing on students engagement and instructor preferences.

\subsection{Methods}

\noindent \textbf{Data source and Demographics.} We gathered data from a pre-course surveys and log traces from across two semesters. The survey collected students' self-reported gender and prior programming experience on a 5-point Likert scale, with the highest level of experience represented by a score of 5. Log traces were used to identify learners' first choice of instructor for interactive contents (code walkthroughs or explanation videos). The study included 662 students from the 2021 semester and 559 students from the 2022 semester who had sufficient exposure to both instructors. Table.~\ref{tab:freq} provides a breakdown of student demographics. Students with minimal participation were excluded.
\begin{table} [h]
\setlength{\belowcaptionskip}{-5pt}
\setlength\tabcolsep{1pt} 
  \caption{Self-reported characteristic data}
  \label{tab:freq}
  \begin{tabular}{llcc}
    \toprule
    \textbf{Category} & \textbf{Subcategory} & \textbf{2021 Fall} & \textbf{2022 Fall} \\
    \midrule
    Gender & Female & 177 & 182 \\
    & Male & 471 & 349 \\
    & Other & 7 & 16 \\
    & Missing & 8 & 12 \\
    \midrule
    Prior experience & 1 & 98 & 92 \\
    & 2 & 268 & 193 \\
    & 3 & 209 & 169 \\
    & 4 & 58 & 66 \\
    & 5 & 22 & 33 \\
    & Missing & 7 & 6 \\
    \midrule
    \textbf{Total} & & 662 & 559 \\
    \bottomrule
  \end{tabular}
  \vspace{-10pt}
\end{table}

\noindent \textbf{Preference Scores and Patterns.} Log traces were used to identify learners' first-choice instructor when viewing interactive content, focusing only on video preferences between two instructors (excluding teaching assistants). We quantified students' preference for a specific instructor based on the frequency of students selecting a video different from the randomly assigned one. Specifically, it is calculated as the difference between the proportion of records selecting instructor A when assigned to instructor B and vice versa, as shown in Equation 1, where n(A|B) represents the total number of times that students watched videos delivered by instructor A when the default video was instructor B and n(B|B) represents the number of times that students selected videos recorded by instructor B while assigned to B. A score of 1 indicated exclusive preference for Instructor A, -1 for Instructor B, and 0 for no preference.  This score was calculated for individual students over weekly lessons, enabling the tracking of preference change and development over the semester. \begin{equation}
\frac{n(A|B)}{n(B|B)+n(A|B)} - \frac{n(B|A)}{n(A|A)+n(B|A)}
\end{equation}

Students were further classified into distinct preference patterns based on scores: favoring a certain instructor ($|\text{preference score}| \geq 0.2$) and having no preference ($|\text{preference score}| < 0.2$) as shown in Table.~\ref{tab:prefpattern}. This threshold was empirically determined by examining the preference score distribution and selecting the one with the largest gap. To check the robustness of this threshold, we also tested thresholds of 0.15 and 0.25. All three thresholds were comparable in the magnitude of statistical comparisons.

\noindent \textbf{Evaluation Approach.} Our study employed a multi-faceted approach to investigate the presence, development, and influencing factors of student instructor preferences in an online CS1 course. Our analysis proceeded as follows: (1) Existence of Pre-existing Preferences: To understand if students came to the course with initial preferences, we examined the distribution of preference scores calculated from the first 20\% of students' video-watching records. This analysis aimed to capture any pre-existing biases or preferences students might have had early in the course. (2) Evolution of Preferences: We analyzed the trajectory of preference scores over the entire semester and conducted a comparative analysis between initial preferences and semester-long preferences. This longitudinal approach allowed us to observe how preferences developed or changed over time. (3) Influence of Student Characteristics: To explore potential factors shaping instructor preferences, we employed Kruskal-Wallis tests to compare continuous preference scores across different student subgroups. Additionally, we used chi-square tests to examine the relationship between categorical preference patterns and student characteristics, particularly focusing on gender and prior programming experience. This analysis aimed to uncover any significant associations between student demographics and instructor preferences.

\subsection{Main Findings}

Our analysis of student preferences for instructors in the online CS1 course revealed several key insights. These findings shed light on the existence of pre-existing preferences, the evolution of preferences over time, and the influence of student characteristics such as gender and prior programming experience. The following subsections detail these results, providing a comprehensive picture of how students interact with and choose between multiple instructors in an online learning environment.

\noindent \textbf{Existence of Pre-existing Instructor Preference.} \textit{Initial preference}, defined based on the first 20\% of students’ instructional video-watching records, was examined to capture preexisting preferences prior to developing familiarity with the instructors. More than half of the students (56.5\%) demonstrated initial preferences, that is, their absolute preference score was equal to or greater than our predefined threshold of 0.2. 

\begin{table} [htb]
\setlength\belowcaptionskip{0pt}
\setlength\tabcolsep{2pt} 
\centering
\caption{Frequency distribution of initial and semester preference patterns}
\label{tab:prefpattern}
\begin{tabular}{lccc}
\toprule
 & \textbf{NoPref} & \textbf{PrefA} & \textbf{PrefB} \\
\midrule
Preference score & (-0.2, 0.2) & [0.2, 1] & [-1, -0.2] \\
Initial preference & 235 (43.5\%) & 211 (39.1\%) & 94 (17.4\%) \\
Semester preference & 351 (53.0\%) & 139 (21.0\%) & 172 (26.0\%) \\
\bottomrule
\end{tabular}
\end{table}

\begin{figure}
\vspace{0pt}
  \centering
  \includegraphics[width=\linewidth]{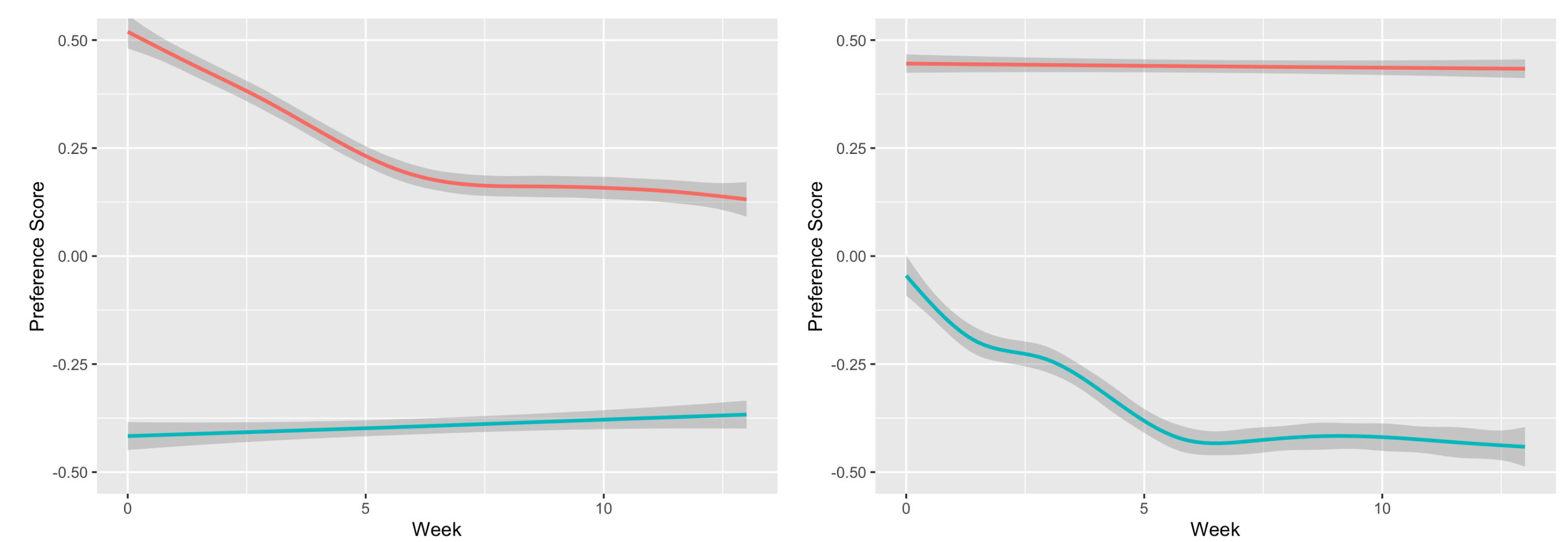}
  \caption{Preference trajectory of students expressing initial preference (left) and semester preference (right). Red lines denote preference for Instructor A; Blue lines denote preference for Instructor B.}
  \label{fig:semester}
  \vspace{-10pt}
\end{figure}

\noindent  \textbf{Consistency of Instructor Preference.} The consistency of the preference was examined by plotting the weekly preference score trajectory of students demonstrating a preference for a particular instructor in the early stage, as well as those demonstrating a preference across the whole semester. Fig. ~\ref{fig:semester} displays the changes in preference scores of students with a pre-existing preference towards one of the two instructors. We found a decreasing trend for the student group with initial PrefA pattern. Students coming with an initial preference for instructor A gradually developed a different preference. The student group with PrefB pattern, instead, had a flatter trajectory shown in Fig. ~\ref{fig:semester}, indicating a more stable preference. 

As a complementary analysis, distributions of preference scores across the entire semester were further examined. Here, we introduced the term semester preference, which accounted for all video-watching logs collected throughout the entire semester. Table~\ref{tab:prefpattern} describes the frequency distribution of semester preference patterns. Fig.~\ref{fig:semester} displays the evolution of preference scores of students who demonstrated certain preference across the semester. It is important to note that the student populations expressing preference in the initial stage may not be identical to those expressing preference throughout the semester. As a matter of fact, descriptive data showed that only 73 out of 211 (35\%) learners with PrefA pattern in the early stage still held the same preference (i.e., semester preference score $\geq$ 0.2) at the end of the semester, indicating a shift in preference. In contrast, 64 out of 94 (68\%) of students with PrefB pattern in the early stage of the course still held a preference for instructor B (i.e., semester preference score $\leq$ 0.2) at the end of the semester. Moreover, 148 out of 235 (63\%) of the students without any initial preference eventually developed a preference for either instructor A or B. This descriptive analysis offers further evidence that not only do students express an initial instructor preference, but these preferences may change throughout the semester.

\noindent  \textbf{Gender Effects on Preference.} Next, we explored the relationship between preference (both initial preference and semester preference) and two students characteristics: gender and self-perceived prior experience on programming. 

\begin{figure} [h]
\vspace{-5pt}
  \centering
  \includegraphics[width=\linewidth]{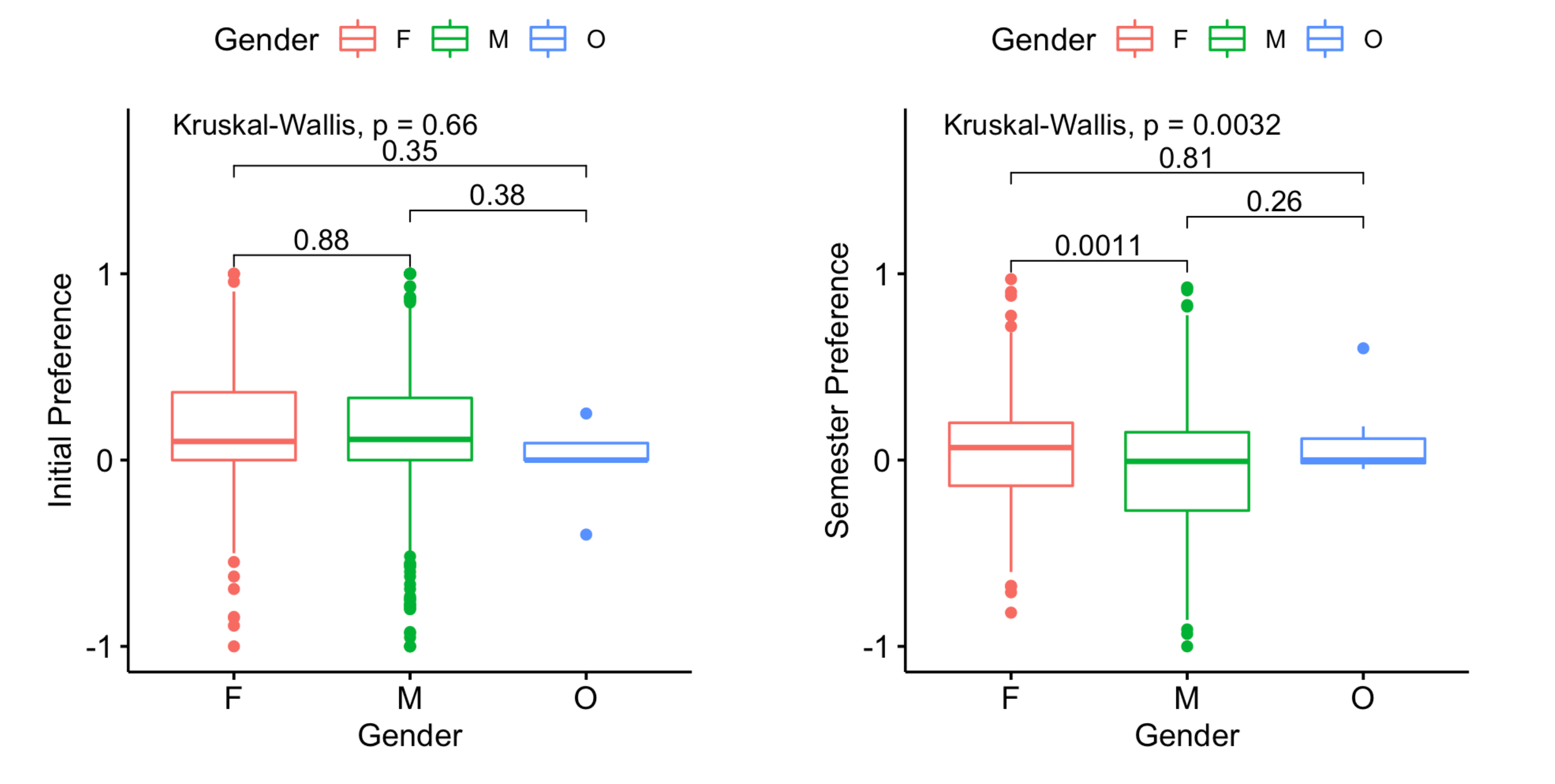}
  \caption{Comparison of the initial preference (left) and semester preference scores (right) with regard to gender in the Fall 2021 semester.}
  \label{fig:gendercompare}
  \vspace{-5pt}
\end{figure}

We first investigated the relationship between gender and preference score. We conducted Kruskal-Wallis tests as the Shapiro-Wilk normality test indicated that the distribution of the preference scores was significantly divergent from a normal distribution. The comparison of the initial preference across different gender groups did not yield significant differences ($H(2) = 0.85$, $p = .655$, $\eta^2 = -.002$) as shown in Fig.~\ref{fig:gendercompare}. However, comparison of the semester preference scores found a significant difference across gender groups ($H(2) = 11.48$, $p = .003$, $\eta^2 = .015$). The effect size was small ($\eta^2 = .015$), based on a common interpretation of Eta squared~\cite{tomczak2014need}, where $0.01$ indicates a small effect, $0.06$ indicates a medium effect, and $0.14$ indicates a large effect. Wilcoxon's pairwise tests revealed that female learners, on average, demonstrated a stronger preference towards instructor A (see Fig.~\ref{fig:gendercompare}) who self-identified as female. This difference remained significant after adjusting the $p$-value to adjust the false discovery rate by means of the Bonferroni correction ($p = .003$).

To complement the analysis of preference scores, we also examined three discrete preference patterns (i.e., female instructor A, male instructor B, and no preference) and examined their frequency distribution across gender subgroups. This approach is widely used in clinical research to facilitate the interpretation and presentation of results \cite{altman2006cost}. Two primary motivations led to the implementation of this strategy: (1) to enhance the readability and clarity of results presented in a tabular format and (2) to provide an alternative and more informative perspective on the distribution of the three preference patterns, rather than focusing on the numerical variations in the preference scores.

A Chi-square test was performed to investigate the relationship between students' self-reported gender and preference patterns. Table~\ref{tab:chi} shows that the results were consistent with those of the Kruskal-Wallis test and revealed a significant association between student gender and their preference patterns over the semester ($\chi^2$(4) = 17.31, $p$ = .002) with a small effect size (.06) as indicated by a Cramer's $V$ statistic. The comparison of observed and expected frequencies suggested that on average female students tended to favor the female instructor A and male students favored male instructor B. Frequency distribution of preference patterns revealed a higher proportion of female students (25.1\%) who preferred the female instructor compared to the male instructor (19.0\%), while male students demonstrated an opposite trend, with a larger proportion preferring the male instructor B (29.4\%) compared to the female instructor A (18.9\%).

\begin{figure} [h]
\vspace{-5pt}
  \centering
  \includegraphics[width=\linewidth]{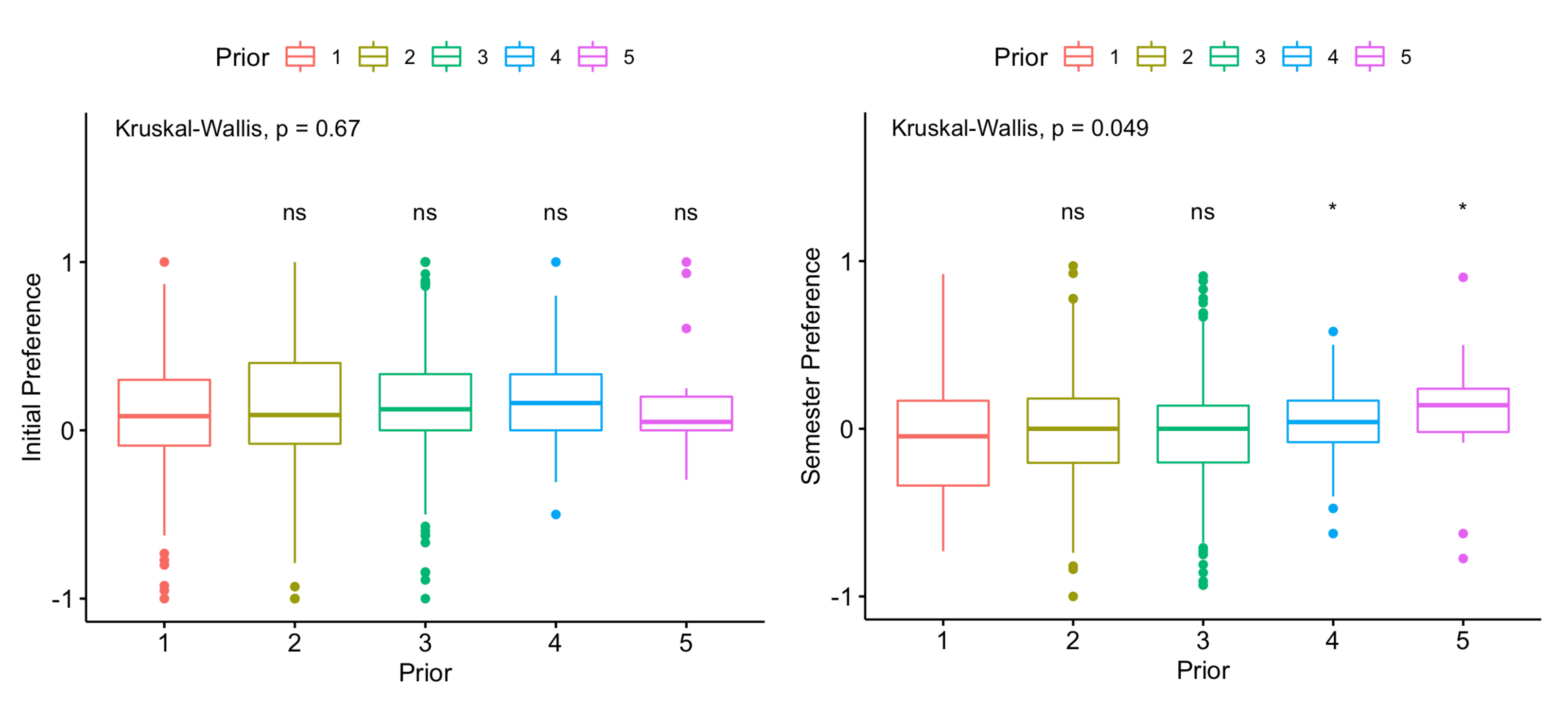}
  \caption{Comparison of the initial preference (left) and semester preference scores (right) with regard to prior programming experience in the Fall 2021 semester.}
  \label{fig:chiresult}
  \vspace{-5pt}
\end{figure}

\noindent  \textbf{Prior Programming Experience Effects on Preference.} Comparing the initial preference scores across learners with different levels of prior experience, the Kruskal-Wallis test yielded no significant difference ($H$(2) = 2.38, $p$ = .667, $\eta^2$ = -.003) as shown in Fig.~\ref{fig:chiresult}. In terms of semester preference scores, significant differences emerged ($H$(2) = 9.53, $p$ = .049, $\eta^2$ = .009) with a small effect size ($\eta^2 = .009$). Wilcoxon's pairwise tests revealed a statistically significant difference between the semester preference score of learners with a rating of 4 ($p$ = .03) or 5 ($p$ = .03) in terms of prior programming experience and those with a rating of 1. These differences were still significant after Bonferroni correction was applied to adjust the $p$-value. Learners who perceived themselves with higher prior experience tended to indicate a preference for instructor A.

Categorizing preference scores into discrete preference patterns found a significant difference among learners with various levels of prior experience in the semester preference patterns but not in the initial preference patterns. Table~\ref{tab:chi} shows that learners with lower and higher levels of prior experience demonstrated opposite trends with regard to their instructor preference.

\begin{table*}[ht]
\vspace{-5pt}
\centering
\caption{Comparison of initial preference and semester preference pattern between learners in different groups}
\label{tab:chi}
\begin{tabular}{lccccccccccc}
\toprule
       & \multicolumn{5}{c}{\textbf{Initial Preference Pattern}} & \multicolumn{5}{c}{\textbf{Semester Preference Pattern}} \\ 
\cmidrule(lr){2-6} \cmidrule(lr){7-12}
       & PrefA & & PrefB & NoPref & \textit{Statistics} && PrefA & & PrefB & NoPref & \textit{Statistics} \\ 
\midrule
Female & 57\textsuperscript{a}(58.4\textsuperscript{b}) && 21(25.7) & 71(64.9) & $\chi^2$(4) = 2.94, \textit{p}\textsuperscript{c} = .568 && 47(36.5) && 29(46.0) & 101(94.5) & $\chi^2$(4) = 17.31, \textit{p} = .002 \\
Male   & 151(148.6) && 70(65.4) & 158(164.9) & && 87(97.0) && 141(122.2) & 242(250.8) & \\
Other  & 1(2.0) && 1(0.8) & 3(2.2) & && 1(1.4) && 0(1.8) & 6(3.7) & \\ 
\midrule % This will create a horizontal line to separate the data
Exp 1 & 29(32.0) && 18(13.8) & 34(35.3) & $\chi^2$(8) = 10.25, \textit{p} = .248 && 24(21.1) && 47(37.3) & 27(39.6) & $\chi^2$(8) = 18.89, \textit{p} = .015 \\
Exp 2 & 86(87.9) && 44(37.9) & 93(97.1) & && 54(56.3) && 102(99.3) & 105(105.5) & \\
Exp 3 & 71(66.7) && 25(28.7) & 73(73.6) & && 36(43.1) && 73(76.1) & 91(80.8) & \\
Exp 4 & 20(17.7) && 2(7.7) & 23(19.6) & && 13(11.0) && 15(19.4) & 23(20.6) & \\
Exp 5 & 5(6.7) && 2(2.9) & 10(7.4) & && 9(4.5) && 3(8.0) & 9(8.5) & \\ 
\bottomrule
\end{tabular}
\textsuperscript{a}An observed frequency. \textsuperscript{b}Expected frequency based on the null hypothesis. \textsuperscript{c}\textit{p}-value is from Chi-square test.
\vspace{-10pt}
\end{table*}

\section{Discussion}
% GC: This first graf seems like filler that could be dropped or condensed
%
Our study explores whether and how students express their preferences in a learning environment where they have repeated opportunities to interact with different instructors through recorded instructional content.
The results reveal several interesting insights.
First, instructor preferences in this online CS1 course were not static but evolved throughout the semester.
Comparing preference patterns identified in the early stage and over the whole semester, we found that students who came to class with a preexisting preference adjusted it at different stages, while those without an initial preference may develop a certain preference by the end of the semester.
Many potential reasons could account for this dynamic, including increased familiarity with both instructors, the development of preferences for one instructor over another, or adjustments based on varying content difficulty.
This dynamic nature of preferences suggests that providing multiple instructor options throughout the course is valuable, as it allows students to adapt their learning experience to their changing needs and preferences.

Furthermore, we observed the emergence of gender-based preferences in the latter half of the semester.
While initial preferences showed no significant gender differences, male and female students tended to favor instructors of their own gender as the course progressed.
Male students, in particular, demonstrated a stronger and consistent preference for male instructors. These findings align with prior research on gender-based preferences in STEM education \cite{macnell2015s, mcgoldrick2002instructor}.
Students may relate more easily to instructors of the same gender, perceiving them as potential role models in the field. 
However, this result should be interpreted with caution due to potential confounding factors, such as pedagogical style and pacing differences between instructors, which are hard to measure explicitly and were not considered in the current study.
Future research should address these limitations by incorporating measures of pedagogical style, expanding to multiple courses and institutions, and investigating how preferences impact learning outcomes.
Additionally, qualitative methods could provide deeper insights into the reasons behind students' instructor preferences.

These findings have several practical implications for institutions and online learning environment design.
First, institutions should consider offering students greater flexibility in selecting instructors.
In larger courses, it is common to divide students into sections, each independently taught by a single instructor. Multi-instructor designs, as demonstrated in our study, might offer a more flexible and personalized learning experience by allowing students to switch between instructors based on their evolving preferences..
This approach, while promising, requires careful coordination to ensure consistency in content delivery across instructors.
Online learning platforms can help multiple instructors collaboratively make additive contributions to a single course, rather than working separately and independently.
Second, given the observed gender-based preferences, institutions should encourage diversity of instructors.
Providing students with the opportunity to learn from instructors with diverse backgrounds can better meet the varied needs and preferences.
Offering this flexibility in instructors not only enhances student satisfaction but also contribute to broader goals of supporting underrepresented groups in STEM fields.

For educators and institutions considering implementing a similar multi-instructor platform, we recommend a gradual approach, starting with a subset of course topics and gathering feedback from both students and instructors before full-scale implementation.
This phased approach will allow for the refinement of coordination mechanisms and content guidelines, ensuring a consistent and high-quality learning experience as the platform scales.

\section{Conclusion and future work}
In summary, this investigation examines the presence and development of student preferences for instructors within an online CS1 course. Utilizing a novel CS1 learning platform that provides access to instructional videos created by different instructors, we observed the presence and development of instructor preference. The statistical analysis uncovers the relationship between gender and the development of instructor preferences. These insights underscore the necessity of incorporating diverse instructional voices and pedagogical approaches to accommodate the distinct needs and expectations of students. Our findings have practical implications for both educational institutions and instructors. Institutions should promote diversity in instructor profiles and support students to gain a more comprehensive understanding of instructors during course selection. Instructors can engage in collaborative teaching to meet the diverse needs, preferences, and goals of students. 

While this study presents a novel perspective on understanding students’ preferences towards instructors, several limitations should be acknowledged. First, this study only examined students’ instructor preferences through recorded instructional videos on an asynchronous learning platform. Future research should consider investigating other learning behaviors reflecting preferences and their impact on learning motivation and outcomes. Second, the context-specific nature of our research may limit its generalizability regarding the preference for specific inspectors. Extending this study across different settings, student populations and incorporating more diverse instructor profiles can provide more comprehensive insights. Third, future research should consider features that will allow students to provide more detailed feedback on why they prefer certain instructors for specific topics, which will provide valuable insights for further improving our platform and instructional approaches.

\section{Acknowledgement}
This study was supported by the National Science Foundation. Any conclusions expressed in this material do not necessarily reflect the views of the NSF.\clearpage

%%\section{Acknowledgement}
%%\input{main/7-acknowledgement}

\bibliographystyle{ACM-Reference-Format}
\balance
\bibliography{main/ref}
%--------------------------------
\end{document}